\documentclass[aps,prl,twocolumn,showpacs,superscriptaddress,groupedaddress]{revtex4}  

\usepackage{epsfig}
\usepackage{longtable}
\usepackage{graphicx}  
\usepackage{dcolumn}   
\usepackage{bm}        
\usepackage{amssymb}   

\def\beq{\begin{equation}}
\def\eeq{\end{equation}}
\def\be{\begin{equation}}
\def\ee{\end{equation}}
\def\bea{\begin{eqnarray}}
\def\eea{\end{eqnarray}}

\def\ms{\overline{\rm MS}}

\def\t{\tilde }

\def\ds{\displaystyle}
\def\sumint{\hbox{$\sum$}\!\!\!\!\!\!\!\int}
 
\newcommand{\lsim}{\raisebox{-0.13cm}{~\shortstack{$<$ \\[-0.07cm] $\sim$}}~}
\newcommand{\gsim}{\raisebox{-0.13cm}{~\shortstack{$>$ \\[-0.07cm] $\sim$}}~}

\begin{document}

\title{Scale Invariant Resummed Perturbation at Finite Temperatures}

\author{Jean-Lo\"{\i}c Kneur}
\email{jean-loic.kneur@univ-montp2.fr}
\affiliation{Laboratoire Charles Coulomb (L2C), UMR 5221 CNRS-Univ. Montpellier, Montpellier, France}

\author{Marcus B. Pinto}
\email{marcus.benghi@ufsc.br} 
\affiliation{Departamento de F\'{\i}sica, Universidade Federal de Santa
  Catarina, 88040-900 Florian\'{o}polis, Santa Catarina, Brazil}

\begin{abstract}
We use the scalar model with quartic interaction 
to illustrate how a nonperturbative variational technique 
combined with renormalization group (RG) properties efficiently resums perturbative expansions 
in thermal field theories.  The resulting convergence and  
scale dependence of optimized thermodynamical quantities, here illustrated up to two-loop order, 
are drastically improved as compared to 
standard perturbative expansions, as well as to other related 
methods such as the screened perturbation or 
(resummed) hard-thermal-loop perturbation, that miss RG invariance as we explain.
Being very general and easy to implement, our method  is a potential analytical alternative to deal with 
the phase transitions of field theories such as thermal QCD.
\end{abstract}

\maketitle
At sufficiently high temperature or density, one could naively hope that the asymptotic freedom 
property of quantum chromodynamics (QCD) would give a reliable perturbation theory (PT) handle 
on the quark-gluon plasma physics. However, it is well-known that severe infrared divergences unavoidably spoil a
standard PT approach in thermal QCD, and generically also for other thermal field theories, such that 
PT gives poorly convergent and furthermore badly scale-dependent results at successive orders 
(see {\it e.g.} \cite{Trev} for reviews). 
Nowadays the development of powerful computers and numerical 
techniques offer the possibility to solve these nonperturbative problems {\it in silico}, 
employing lattice field theory (LFT). 
So far LFT has been very successful in the description of the QCD phase transitions at 
finite temperatures and near vanishing baryonic 
densities, with results~\cite{aoki} which can be directly used for interpreting the experimental 
output from  heavy ion experiments envisaged to scan over this particular region of the phase diagram.   
However, the well-known numerical sign problem \cite{signpb}, which plagues this method when one considers 
the possibility of a particle-antiparticle asymmetry (signaled by a finite chemical potential),  
prevents LFT to be successfully used to describe 
compressed baryonic matter. Therefore, at the present stage, one cannot rely on LFT to describe the physics 
of compact stellar objects nor to explore the complete QCD phase diagram. 
In parallel over the last decades many efforts have been devoted to try to understand more analytically 
the bad convergence generically observed for thermal PT, even for moderate coupling values.
Typically the dynamical generation of a 
thermal screening mass $m_D\sim \sqrt{\lambda} T$ influences 
the relevant expansion of thermodynamical quantities, such as the pressure, involving 
powers of $\sqrt \lambda$ rather than only $\lambda$. Accordingly the predictions are,
a priori, less convergent than for the $T=0$ case. A plethora of 
nonperturbative approximations attempting to resum thermal perturbative expansions 
have been developed and refined over the years\cite{Trev,SPTearly,NPRG,Blaizot-Wschebor}. 
The so-called optimized perturbation theory (OPT) is a variational approach 
in which a related solvable case is rewritten in terms of an unphysical parameter,  
allowing for optimized nonperturbative results. In the past decades this strategy has been recycled, 
appearing under different names~\cite{OPT-LDE,odm,pms}. 
 At each successive order of such a modified perturbative expansion, the 
arbitrary variational mass is fixed by a stationary condition. 
This strategy has already been used in a variety of 
different physical situations, including {\it e.g.} the determination of the critical temperature for homogeneous 
Bose gases~\cite{bec2,beccrit}, 
the phase diagram of magnetized planar fermionic systems\cite{optGN,tulio}, and the evaluation of 
quark susceptibilities within effective 
QCD inspired models\cite{njlprc}. The development of a similar method, 
known as screened perturbation theory (SPT)\cite{SPT} or its version tailored 
to treat thermal gauge theories\cite{HTL}, 
hard-thermal-loop (resummed) perturbation theory (HTLpt)\cite{htlpt1},  
have been pushed to three-loop perturbative order~\cite{SPT3l,OPT3l,HTLPT3loop,HTLPTMU}. Given the inherent technical  
difficulties of the (three loop) evaluation of the QCD pressure for the case of hot and dense 
quark matter, the recent results in \cite{HTLPTMU} represent an impressive achievement. Moreover  
their agreement with LFT simulations is quite remarkable 
down to about twice the critical temperature, for the 
scale choice $\mu=2\pi T$ in the $\ms$ renormalization scheme. However, the SPT/HTLpt
presents several shortcoming overshadowing
its potential as a reliable nonperturbative alternative to
LFT. Perhaps the most embarrassing issue is the strongly {\em enhanced} 
scale dependence displayed at increasing two- and three-loop orders, at odds with intuitive expectations: 
at three-loop order even moderate scale variations dramatically affect thermodynamical quantities by 
relative ${\cal O}(1)$ variations~\cite{SPT3l,HTLPT3loop,HTLPTMU}. Another issue with the standard 
variational methods such as OPT, SPT or HTLpt is that beyond lowest orders, optimization gives more and more 
solutions, with unphysical complex-valued ones, often leading to use alternative prescriptions 
like replacing the variational mass with a purely perturbative mass~\cite{HTLPT3loop}, therefore loosing 
valuable nonperturbative information.\\
Recently, the OPT method at vanishing temperatures and 
densities has been consistently 
combined with renormalization group (RG) properties\cite{rgopt1,rgopt_Lam,rgopt_alphas}. 
The resulting RGOPT gives stable and precise results for the Gross-Neveu 
mass-gap~\cite{rgopt1}, and new independent 
determinations~\cite{rgopt_alphas}
of the basic QCD scale ($\Lambda_{\overline {\rm MS}}$) and related coupling $\alpha_S$, or the
quark condensate~\cite{rgopt_qq}. Moreover, often unique and real optimization solutions 
can be obtained~\cite{rgopt_alphas}, by matching those 
solutions to the RG behavior for small couplings; and by using appropriate renormalization scheme changes. \\
Here, we take an important step forward, by showing that the RGOPT is also compatible with the introduction 
of control parameters such as the temperature. To illustrate how to implement the procedure 
we have chosen a simple, yet versatile model so that one 
can easily grasp the basic ideas and follow the main steps when performing 
a particular application. More detailed results and formulas are given elsewhere~\cite{inprepa}.
Aside from purely calculation difficulties, the method described in this Letter 
can be directly extended to a large class of models. 

\begin{figure}[h]
\epsfig{figure=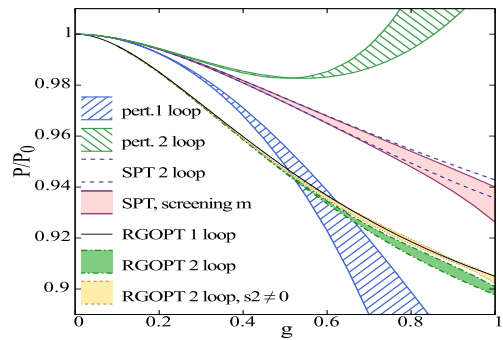,width=6.5cm,height=4.4cm}
\caption[long]{RGOPT $P/P_0(g\equiv \sqrt{\lambda/24})$ at 
one- and two-loop versus standard perturbative and two-loop SPT pressures
 with scale dependence $\pi T <\mu <4\pi T$. }
\label{Prgopt1and2loop}
\end{figure}

We thus start by considering the Lagrangian for one neutral scalar field with a quartic interaction,
\be
{\cal L} = \frac{1}{2} \partial_\mu \phi \partial^\mu \phi -\frac{m^2}{2}\phi^2 -\frac{\lambda}{4!} \phi^4\,\,,
\label{Lphi4}
\ee
where we have introduced a generic mass term $m$. 
The textbook result for the two loop free energy (equivalently minus the pressure) is \cite{kapusta,SPT3l}
\be
{\cal F}_0 = \frac{T}{2}
\sumint_p \ln (\omega_n^2+\omega_{\bf p}^2)  +
\frac{\lambda T^2}{8} \left(\sumint_p \frac{1}{\omega_n^2+\omega_{\bf p}^2} \right)^2 +{\cal F}_0^{\rm ct},
\label{2lbasic}
\ee
where in the imaginary time formalism $\omega_n=2\pi T n$ $(n=0,1,\cdots)$ 
represents the bosonic Matsubara frequencies and $\omega_{\bf p}^2 ={\bf p}^2 +m^2$ is the dispersion relation. 
The sum-integral in (\ref{2lbasic}) as usual represents the sum over Matsubara frequencies and remaining 
integration with measure $d^{3-2\epsilon} {\bf p}/(2\pi)^{3-2\epsilon}$
using dimensional regularization to perform the integral. The one-loop part of (\ref{2lbasic}) is
\be
(4\pi)^2 {\cal F}_0 = -\frac{m^4}{8} \left[\frac{2}{\epsilon}+3+2\ln (\frac{\mu^2}{m^2} ) \right]
 +{\cal F}_0(T)  +{\cal F}_0^{\rm ct},
\label{1lbasic}
\ee
where $\mu$ is the arbitrary renormalization scale in the $\overline {\rm MS}$ 
renormalization scheme, and ${\cal F}_0^{\rm ct}=m^4/(4\epsilon)$
represents the vacuum energy counterterm\cite{SPT3l}. 
We can already address a crucial point by considering the one-loop part free energy (\ref{1lbasic}).
It is a trivial matter to check that the renormalized result spoils perturbative RG invariance. 
Acting on Eq.~(\ref{1lbasic}) with the standard RG operator:
\be
\mu\frac{d}{d\,\mu} =
\mu\frac{\partial}{\partial\mu}+\beta(\lambda)\frac{\partial}{\partial \lambda}
+\gamma_m(\lambda)\,m
 \frac{\partial}{\partial m}\;,
 \label{RGop}
\ee 
and noting that the thermal contribution ${\cal F}_0(T)$ is scale independent,  
yields a 
remnant contribution: $-m^4/2$, not compensated
by lowest orders terms from $\beta(\lambda)$ or 
$\gamma_m(\lambda)$ in (\ref{RGop}),  
those being at least of next order ${\cal O}(\lambda)$. This is a manifestation of the fact that perturbative 
RG invariance generally occurs from cancellations between terms from the RG equation 
at order $\lambda^k$ and the explicit $\mu$ dependence at the next order $\lambda^{k+1}$   
(our normalization is $\beta(\lambda)\equiv d\lambda/d\ln\mu= b_0 \lambda^2 
+b_1 \lambda^3 +\cdots$ for the $\beta$ function and 
$\gamma_m(\lambda) \equiv d \ln m/d\ln\mu= \gamma_0 \lambda +\gamma_1 \lambda^2 +\cdots$
for the anomalous mass dimension,
with~\cite{RGphi4loop} $(4\pi)^2 b_0 = 3;\; (4\pi)^2 \gamma_0 = 1/2;\; (4\pi)^4 b_1 = -17/3;\;(4\pi)^4 \gamma_1 = -5/12$). 
Nevertheless, perturbative RG invariance can easily be restored by adding a 
finite vacuum energy term, ${\cal E}_0$, to the action without changing the dynamics. 
Although this term is usually ignored, minimally set to zero in the (thermal) literature~\cite{SPT3l,htlpt1,HTLPT3loop}, 
we stress that it is instrumental for perturbative RG invariance to be achieved. 
Not surprisingly, we claim it largely explains the degrading scale-dependence 
at higher orders in other similar resummation methods like SPT and HTLpt, 
which ignore those finite vacuum energy terms. 
The subtraction in $\ms$-scheme is conveniently written as\cite{qcd2,rgopt_alphas,rgopt_qq}:
\be
{\rm {\cal E}_0}(\lambda,m) = -(m^4/\lambda)\: \sum_{k\ge 0} s_k  \lambda^{k},
\label{subgen}
\ee
where the coefficients $s_k$ are 
perturbatively determined order by order from RG invariance. In the normalization of Eq.~(\ref{1lbasic}) we find 
$s_0 =[2(b_0-4\gamma_0)]^{-1} = 8\pi^2$, so that when augmented with ${\cal E}_0$ 
the renormalized free energy from Eq.~(\ref{1lbasic}) is RG invariant at the one loop level. 
This can be carried out to higher orders, to give  
$s_1 =-1$, $s_2=(23+36\zeta[3])/(480\pi^2)$, etc. 
Note that the apparently singular behavior  
for $\lambda \to 0$ in (\ref{subgen}) will actually disappear from the final optimized free energy. 
We stress that the previous construction, being only dependent on the renormalization procedure, 
does not depend on temperature-dependent 
parts: at arbitrary perturbative orders the $s_k$ coefficients can be determined 
from the $T=0$ contributions only. This is indeed well-known, and  
the non-RG invariant remnant part defines the
so-called vacuum energy anomalous dimension, that has been calculated even to five-loop
order for the general $O(N)$ scalar model~\cite{vacanom_kastening}. 
Our independent results for the $s_k$ are fully consistent with \cite{vacanom_kastening}. 
A subtlety is that according to Eq.~(\ref{subgen}), $s_k$ is strictly required for
RG invariance at order $\lambda^k$, but contributes at order $\lambda^{k-1}$.
So at order $\lambda^k$ one may minimally choose to include only $s_0, \cdots s_k$, or  
more completely include $s_{k+1}\ne 0$, thus incorporating higher order RG information.\\ 
One can now proceed to apply the RGOPT resummation, by first performing on the RG-invariant free energy 
the substitution which appropriately modifies its perturbative expansion:
\be m^2 \to m^2 \,(1-\delta)^{2a}\,;\;\;\lambda \to \delta \lambda, 
\label{subst1}
\ee
where now $m$ is an arbitrary parameter, and the role of $a$ is explained below. 
One then re-expands at successive orders $\delta^k$, setting $\delta\to 1$ in the final results.
This procedure is consistent with renormalizability~\cite{gn2,chiku,optON} and gauge 
invariance~\cite{qcd2}, whenever the latter is relevant. The arbitrary mass parameter $m$ is then most 
conveniently fixed by a variational optimization prescription~\cite{pms}:
\be 
\frac{\partial{\cal F}_0^{(k)}}{\partial m}(m,\lambda,\delta=1)|_{m\equiv \t m}  \equiv 0,
\label{OPT}
\ee
and $\t m\ne 0$ determines a nontrivial mass $\t m(\lambda)$ with nonperturbative $\lambda$-dependence. \\
In most previous OPT~\cite{OPT-LDE} (similarly SPT\cite{SPT} and HTLpt\cite{HTL}) applications, 
the linear $\delta$-expansion was used, {\it i.e.} assuming $a=1/2$
in (\ref{subst1}) mainly for simplicity and economy of parameters. 
However, to preserve RG invariance after 
performing (\ref{subst1}), $a$ is uniquely fixed~\cite{rgopt_alphas} by the
universal (renormalization scheme independent) first order RG coefficients, as we recall below. 
Note, once combined with Eq.~(\ref{OPT}), the RG Eq.~(\ref{RGop}) takes a reduced {\em massless} form
\be
\left[\mu\frac{\partial}{\partial\mu}
+\beta(\lambda)\frac{\partial}{\partial \lambda}\right]{\cal F}_0^{(k)}(m,\lambda,\delta=1)=0,
\label{RGred}
\ee
so Eq.~(\ref{RGred}) with the OPT Eq.~(\ref{OPT}) completely set  
{\em optimized} $m\equiv \t m$ and $g\equiv \t g$ ``variational fixed-point" values.\\
Consider the one-loop Eq.~(\ref{1lbasic}), at $T=0$, augmented 
by ${\cal E}_0 = -(m^4/\lambda)s_0$, where as discussed above, $s_0=8\pi^2$. 
Performing (\ref{subst1}), expanding to order $\delta^0$ consistently, 
and taking {\em afterwards} $\delta\to 1$ yields
\be
(4\pi)^2 {\cal F}^{\delta^0}_0 = 
m^4 \left[-\frac{s_0}{\lambda}(1-4a) -\left (\frac{3}{8}+\frac{1}{4}\ln\frac{\mu^2}{m^2} \right )\right].
\label{F0opt0}
\ee
Then to satisfy Eq.~(\ref{RGred}) implies 
$a=\gamma_0/b_0=1/6$. At this one-loop order the RG Eq.~(\ref{RGred})
gives no further constraints, but at higher orders it fixes an optimized coupling, and 
$a=\gamma_0/b_0$ guarantees that among both RG and OPT solutions, at least one (often unique) 
is consistent with the $T=0$ standard perturbative behavior~\cite{rgopt_alphas} for $\lambda\to 0$,
{\it i.e.} infrared freedom in the present case: $\lambda(\mu\ll m)\simeq
[b_0 \ln (m/\mu)]^{-1}$. \\
Switching on thermal effects, it is convenient
to express our results in terms of the one-loop renormalized self-energy including all $T$-dependence,  
$\Sigma_R$, explicitly\cite{Trev,SPT3l} 
\be
\Sigma_R 
=\gamma_0 \lambda \left[m^2 \left (\ln\frac{m^2}{\mu^2} -1\right )+T^2 J_1\left (\frac{m}{T}\right )\right],
\label{Sigma}
\ee
with the thermal integrals ($t=p/T$ and $x=m/T$):
\be
J_n(x) = \frac{4\Gamma[1/2]}{\Gamma[5/2-n]} \,\int_0^\infty dt \frac{t^{4-2n}}{\sqrt{t^2+x^2}}\,
\frac{1}{e^{\sqrt{t^2+x^2}}-1} .
\label{Jn}
\ee
Then noting that $T\frac{\partial}{\partial m^2} \sumint\, \ln (\omega_n^2+\omega_{\bf p}^2)=2 
\Sigma_R/\lambda$, the exact solution of the one-loop OPT Eq.~(\ref{OPT}) takes the 
form of a self-consistent ``gap" equation: 
\be
\t m^2 = (4\pi)^2\,b_0\:\Sigma_R (\t m^2),
\label{opt0Tex}
\ee
which is exactly scale-invariant by construction. 
To illustrate this more explicitly, it is convenient to use the high-$T$ expansion $ m/T\equiv x \ll 1 $ 
of $J_n(x)$, {\it e.g.}  $J_0(x) \simeq 16\pi^4/45 -4\pi^2 x^2/3+ 8\pi x^3/3 +x^4 
(\ln x/(4\pi)+\gamma_E -3/4)  +{\cal O}(x^6)$. This approximation is actually valid at
the $0.1\%$ level even for $x\lsim 1$, sufficient for our purpose since the RGOPT one-loop solution
$\t m/T$ always lies in this range.
In this case the OPT Eq.~(\ref{OPT}) is a simple quadratic
equation for $x$, with the unique physical ($x>0$) solution: 
\be
\ds \t x = \frac{\t m^{(1)}}{T} = \pi  \frac{\sqrt{1+\frac{2}{3}\left (\frac{1}{b_0\lambda}+ L_T\right )}-1}
{\frac{1}{b_0\lambda}+ L_T} .
\label{solopt0T}
\ee
 with $L_T\equiv \ln [\mu\,e^{\gamma_E}/(4\pi T)]$.
We stress that the variational mass (\ref{solopt0T}) is 
unrelated to the physical screening mass~\cite{mdeb}.
The corresponding one-loop RGOPT pressure reads
($P_0=(\pi^2/90)T^4$ is the ideal gas pressure):
\be
\frac{P^{(1)}}{P_0} = 1 -\frac{15}{4\pi^2}\t x^2 +\frac{15}{2\pi^3}\t x^3
+ \frac{45}{16\pi^4} \left (\frac{1}{b_0\lambda}+L_T\right ) \t x^4.
\label{P1P0}
\ee
Eqs.~(\ref{opt0Tex})-(\ref{P1P0}) have clearly a nonperturbative dependence in $\lambda$,
and are {\em exactly} scale-invariant, 
upon using for $\lambda\equiv\lambda(\mu)$ the ``exact" (one-loop) running:
$1/\lambda(\mu')=1/\lambda(\mu) -b_0 \ln \mu'/\mu$, then $1/(b_0\lambda(\mu)) + L_T $ is explicitly
$\mu$-independent. Thus Eqs.~(\ref{solopt0T}) and (\ref{P1P0}) only depend on the single 
parameter $b_0\lambda(\mu_0)$, where $\mu_0$ is some reference scale, typically $\mu_0=2\pi T$.
This is a remarkable result, recalling that we started from (\ref{1lbasic}) augmented by  
$-m^4 (s_0/\lambda)$ being RG invariant up to neglected higher order ${\cal O}(\lambda)$, but not yet resummed,
while (\ref{solopt0T}), (\ref{P1P0})
are {\em all-order} RG invariant, showing the resummation efficiency after optimization. 
Eq.~(\ref{P1P0}), perturbatively expanded, gives for the first few orders
$P^{(1)}/P_0 \simeq 1 -5\alpha/4 +5 \sqrt{6} \alpha^{3/2}/3 +5(L_T-6) \alpha^2/4 +{\cal O}(\alpha^{5/2})$
where $\alpha\equiv b_0 \lambda$.\\
Eqs.~(\ref{opt0Tex})-(\ref{P1P0}) reproduce exactly at arbitrary orders 
the $O(N$) scalar model large 
$N$-results ({\it e.g.} Eq.~(5.7) of \cite{phi4N}), as can be checked upon identifying the correct
large-$N$ $b_0 =1/(16\pi^2)$ value~\cite{phi4N}.
These results are also equivalent to those (at {\em two-loop} order) in \cite{2PI}, {\em if}
replacing $b_0=3/(16\pi^2)$ by $b_0/3$, as argued in \cite{2PI}. As we keep the correct $b_0$,  
Eq.~(\ref{P1P0}) differs from standard perturbative pressure by  
$\lambda(\mu_0) \to \lambda_{pert}(\mu_0)/3$: this is not a problem, simply a different calibration 
as $\lambda$ is yet arbitrary since the model is not fully specified by any data fixing a physical input scale, $\mu_0$. 
Indeed, this apparent discrepancy disappears if expressing our results in terms of the {\em physical} mass: 
to see it, we solve Eq.~(\ref{OPT}) now for $\t \lambda(m)$, replace it in (\ref{P1P0}), it 
gives simply: $P^{(1)}/P_0 = 1-15x^2/(8\pi^2)+15x^3/(8\pi^3)+{\cal O}(10^{-4}x^6)$. But here   
$x= m/T$ is arbitrary as we already used (\ref{OPT}) to fix $\t \lambda(m)$. Now taking for $m$ the physical screening mass~\cite{mdeb} 
$m^2\simeq (\lambda/24)T^2 (1-\sqrt{6\lambda}/(4\pi)+\cdots)$,
exactly reproduces the first two terms of the standard physical pressure~\cite{Trev}. \\ 
Eq.~(\ref{P1P0}) is plotted in Fig. \ref{Prgopt1and2loop}, compared
with standard perturbative expansions at one- and two-loop orders with their notoriously bad scale 
dependence~\cite{Trev}. Note that at one-loop, including non-minimally $s_1\ne 0$ in (\ref{subgen}) 
is actually equivalent to a simple scale redefinition, $\mu\to \mu\, e^{2s_1} =\mu\, e^{-2}$, in all our results above. \\
The two-loop ${\cal O}(\delta^1)$ contribution to the free energy, for $ \delta= 1$, takes 
a compact form in terms of $\Sigma_R$ in Eq.~(\ref{Sigma}):
\be
{\cal F}_0 =\frac{ {\cal E}^{\delta^1}_0}{(4\pi)^2} + \frac{T}{2}
\sumint_{\bf p} \ln (\omega_n^2+\omega_{\bf p}^2) -\left (\frac{2\gamma_0}{b_0}\right )\frac{m^2}{\lambda}\,\Sigma_R +
\frac{\Sigma_R^2}{2\lambda},
\label{del2compact}
\ee
where ${\cal E}^{\delta^1}_0=-m^4/[1/(3b_0\lambda)+s_1/3]$  from (\ref{subgen}), and 
by abuse of notation the finite part of this 
already renormalized expression is meant. 
The exact two-loop OPT and RG Eqs.~(\ref{OPT}) and (\ref{RGred}) can be written compactly as
\bea
&f_{\rm OPT}= \frac{2}{3} h \left (1-\frac{1}{b_0\lambda} \right ) +\frac{2}{3} S +\Sigma^\prime_R 
\left (S-\frac{1}{3\lambda}\right )\equiv 0,\nonumber \\
&f_{\rm RG}=  h \left[ \frac{1}{6}+\left (\frac{b_1}{3b_0}-S\right )\lambda \right ] + 
\frac{1}{2} \beta^{(2)}(\lambda) S^2 \equiv 0,
\label{RG2l}
\eea
with $h\equiv (4\pi)^{-2}$, $\beta^{(2)}(\lambda)= b_0\lambda^2 +b_1\lambda^3$,
and the reduced (dimensionless)
self-energy $S(m,\mu,T)\equiv \Sigma_R/(m^2\lambda)$.
We also have from Eq.~(\ref{Sigma}): 
$\Sigma^\prime_R\equiv \partial_{m^2}(\Sigma_R)= \lambda (S+m^2 S^\prime) =
\gamma_0\lambda \,[\ln (m^2/\mu^2) -J_2(m/T)]$.
One may also solve the OPT and RG equations in the high-$T$ expansion approximation, which
is excellent up to 
large (rescaled) coupling $g\equiv \sqrt{\lambda/24}\sim {\cal O}(1)$ values and
gives exactly solvable cubic and quartic algebraic equations respectively, with unique physical
solutions ($\t m/T >0$ etc) easily identifiable. The resulting OPT and RG 
solutions for $\t m/T$ and $P/P_0$ are consistent with Eqs.~(\ref{solopt0T}), (\ref{P1P0})
for the first two order terms perturbatively re-expanded, but contain appropriate modifications
at higher orders (detailed expressions are given elsewhere~\cite{inprepa}). \\
The exact two-loop pressure $P/P_0$ obtained from 
the RG Eq.~(\ref{RGop}), as a function of $g\equiv \sqrt{\lambda/24}$, 
is plotted in Fig. \ref{Prgopt1and2loop}, with scale dependence from exact two-loop running, compared 
with one-loop RGOPT and standard perturbative one- and two-loop pressure.
The RGOPT improvement on convergence and scale dependence as compared to standard perturbative results
is drastic, although a moderate residual scale dependence appears at two-loop, visible on the figure
for (rescaled) coupling values $g \gsim 0.6$.  
This is not surprising since the construction relies on
a two-loop truncated basic free energy. At one-loop RGOPT the exact scale invariance is due to the
peculiar form of the exact running coupling perfectly matching (\ref{solopt0T}).
At two-loop RGOPT the residual scale dependence reappears first at order $\lambda^3$:
$\Delta P_{RGOPT}^{(2)} (\mu) \simeq (0.075 \ln \mu/\mu_0 -1.92) g^6$, 
{\it i.e.} one order higher than the normally expected $\lambda^2$ from standard RG 
properties. 
Moreover, including non-minimally $s_2\ne 0$ (thus catching a 
RG part of the three-loop contributions) modifies the perturbative pressure only at order $\lambda^3$, but slightly
improves further the (nonperturbative) scale dependence, as intuitively expected and seen in Fig. \ref{Prgopt1and2loop}.
More remarkably with $s_2\ne 0$ the two-loop pressure almost coincides with the one-loop result up to relatively large
$g\sim 1$. In Fig. \ref{Prgopt1and2loop} we also compare with the OPT/SPT two-loop 
result, {\it i.e.} discarding ${\cal E}_0$ in (\ref{subgen}), taking $a=1/2$ in (\ref{subst1}), and using
Eq.~(\ref{OPT}); and another prescription using instead the screening 
mass\cite{mdeb}, 
similar to the QCD HTLpt prescription~\cite{HTLPT3loop}. 
Note that the missing one-loop RG invariance 
from unmatched $ m^4 \ln\mu$ terms in (\ref{1lbasic}) remains somewhat hidden at 
one- and two-loop thermal expansion order,  since perturbatively $ m^4 \sim \lambda^2 $, explaining why it plainly  
resurfaces at three-loop $\lambda^2$ order in SPT\cite{SPT3l} or similarly HTLpt\cite{HTLPT3loop}. 
In contrast the RGOPT scale dependence should further improve at higher orders: built on 
perturbative RG invariance at order $k$ for {\em arbitrary} $m$, the mass gap will exhibit remnant scale dependence as 
$\t m^2 \sim \lambda T^2 (1+\cdots+{\cal O}(\lambda^{k+1}\ln\mu))$, thus the dominant scale dependence in the free energy, 
coming from the leading term $-s_0\, m^4/\lambda$, should be ${\cal O}(\lambda^{k+2})$.\\
Finally we can combine the OPT and RG Eqs.~(\ref{RG2l}) to obtain the {\em full}
two-loop RGOPT solution, fixing $\t m/T$ {\em and} $\t \lambda=24\t g^2$ for a given input scale $\mu$. We find for 
$\mu=2\pi T$: $\t m/T \simeq 0.912;\;\t g\simeq 0.825;\;
P^{(2)}_{\rm RGOPT}/P_0\simeq 0.907$, 
and the scale variation for $\pi T< \mu < 4\pi T$ is consistent with the one above shown.\\
In conclusion, we have shown   
how resummations of thermal perturbative expansions based on a variational mass should be appropriately
modified to restore perturbative RG invariance, missed by previous OPT/SPT,HTLpt analogous methods. 
The resulting RGOPT has a different 
interpolation prescription, Eq.~(\ref{subst1}), uniquely
dictated by universal first order RG coefficients $a=\gamma_0/b_0$. The 
RG equation gives an alternative constraint 
to determine the nonperturbative variational mass and coupling, instead of 
using solely the optimization (\ref{OPT}). The RGOPT pressure
has exact one-loop RG/scale invariance, and a scale dependence and stability at two-loop order
drastically reduced up to relatively large coupling values as compared with most other resummation approaches. 
For thermal QCD we anticipate a similarly improved scale-dependence and stability
from appropriate RGOPT adaptations of HTLpt.
\acknowledgements 
M.B.P. is  supported by CNPq.\\

\end{document}